\documentclass[bm,aps,showpacs,amsfonts,amssymb]{revtex4}  

\usepackage{graphicx}
\usepackage{natbib}

\usepackage{color}
\newcommand\beqa{\begin{eqnarray}}
\newcommand\eeqa{\end{eqnarray}}
\newcommand\n{\nonumber\\}
\begin{document}

{~}

\title{
New approach to solution generation using $SL(2,R)$-duality of a dimensionally reduced space in five-dimensional minimal supergravity and new black holes 
}
\vspace{2cm}
 \author{Shun'ya Mizoguchi\footnote{E-mail:mizoguch@post.kek.jp} and Shinya Tomizawa\footnote{E-mail:tomizawa@post.kek.jp}}
\vspace{2cm}
\affiliation{
Theory Center, Institute of Particle and Nuclear Studies,
KEK, Tsukuba, Ibaraki, 305-0801, Japan 
}
\begin{abstract} 
The dimensional reduction of (the bosonic sector of) five-dimensional minimal supergravity to four dimensions leads to a theory with a massless axion  
and a dilaton coupled to gravity and two $U(1)$ gauge fields (one of which has Chern-Simons coupling), 
 whose field equations have $SL(2,R)$ invariance. Utilizing this $SL(2,R)$-duality, we provide a new formalism for solution generation. As an example, applying it to the Rasheed solutions, which are known to describe dyonic rotating black holes (from the four-dimensional point of view) of five-dimensional pure gravity, 
we obtain rotating Kaluza-Klein black hole solutions in five-dimensional minimal supergravity. 
We also show that the solutions have six charges: 
mass, angular momentum, Kaluza-Klein  electric/magnetic charges and electric/magnetic charges of the Maxwell field, four of which are related by a constraint.
\end{abstract}

\date{\today}
\maketitle

\section{Introduction}\label{sec:intro}
Black holes are one of the most mysterious objects in Nature. 
While really observed in the universe, black holes have attracted
much attention from many theorists since the first exact solution 
was found in 1916 \cite{Sch} both in general relativity 
and string theory. By the middle 90's, it became recognized 
that extended black objects were quite common solutions (see \cite{p-brane} for a review) 
to equations of motions of effective low-energy {\it super}gravity of string 
theory, and they are understood as macro-scopic realizations of 
D-branes, which reduce to black holes if all the 
extended dimensions wrap around cycles of some compact internal 
manifold. Various duality transformations have been applied to 
various combinations of a stack of D-branes to yield a variety of 
black hole solutions. In some cases this correspondence  
allows to reproduce black hole entropies by counting the number of 
D-brane states \cite{SV}.

\medskip

In most cases in higher dimensions,    
{\sl asymptotically flat} black holes have been considered in various theories; 
the stationary, axisymmetric black holes (with multiple rotational symmetries) 
of this category are simple generalizations of the well-known four-dimensional black holes 
to higher dimensions. Since, however, our real observable world is macroscopically four-dimensional,  extra dimensions 
have to be compactified in realistic spacetime models in a certain appropriate way.
Therefore it is of great interest to consider higher dimensional Kaluza-Klein black holes with compact extra dimensions, which behave as 
a higher-dimensional object near the horizon but look like a four-dimensional one
 for an observer at large distances~\footnote{
When the black holes will be small enough compared with the size of the extra dimensions, they may be well approximated by the higher dimensional asymptotically flat black hole solutions.}.
Because of the lack of global geometrical symmetry,  the construction of Kaluza-Klein black holes is a non-trivial problem. 
Examples of known exact {\sl Kaluza-Klein black holes} are the ones of cohomogeneity one, which are obtained by squashing the same class of non-compactified black hole solutions~\cite{IM,Wang,NIMT,TIMN,TI,T}.

\medskip
Kaluza-Klein black hole solutions which asymptote to 
some non-trivial $S^1$-bundle over  the four-dimensional Minkowski space-time
were also generalized to supersymmetric solutions. The first supersymmetric Kaluza-Klein black hole solution of this type was 
a black hole in Taub-NUT space~\cite{Gauntlett0,Gaiotto}. A similar type of supersymmetric black hole  was obtained in Ref.~\cite{Elvang3} by taking a
black-hole limit of the supersymmetric black ring in Taub-NUT space; Further generalizations to other supergravity theories, or to black strings and 
black rings in Taub-NUT space were also considered by many authors~\cite{BKW,Bena,Bena3,BGRW,Elvang3,EEMR2,FGPS,CEFGS,GRS,CBJV,T,Gibbons-Perry,CY-KK3,CY-KK4,CY-KK5,Nelson}.
In the  non-extremal cases, a similar type of Kaluza-Klein black hole was considered 
by Ishihara-Matsuno~\cite{IM}, who have found static charged Kaluza-Klein black hole solutions in 
five-dimensional Einstein-Maxwell theory by using the squashing technique in the five-dimensional Reissner-Nordstr\"om solution. 
Subsequently the Ishihara-Matsuno solution was generalized to many different cases in 
five-dimensional supergravity theories. 
In Ref. \cite{NIMT}, the squashing transformation was applied 
to the five-dimensional Cveti{\v c}-Youm charged rotating black hole 
solution~\cite{CY96} with equal charges, and 
as a result, a non-extremal charged rotating Kaluza-Klein black hole solution 
in the supersymmetric limit~\cite{Gauntlett0,Gaiotto} was obtained.  
Furthermore, the application of the squashing transformation to non-asymptotically 
flat Kerr-G\"odel black hole solutions~\cite{Herdeiro0,Gimon-Hashimoto,Herdeiro,Wu} was first 
considered in Refs. \cite{TIMN, MINT, TI}. These solutions \cite{NIMT,TIMN,TI} correspond to 
a generalization of the Ishihara-Matsuno solution to the rotating black holes in five-dimensional minimal supergravity.

\medskip
\begin{table}
\begin{center}
\begin{tabular}{l|llllll}\hline
{\sl Solutions in $D=5$ minimal supergravity} & {\ }$M$ & $J$ & $Q$ & $P$ & $q$ & $p$ \\ \hline
{\sl Gaiotto-Strominger-Yin}~\cite{Gaiotto} &\  yes${}^\dagger $ & no & yes & yes${}^\dagger $ & yes${}^\dagger $ & no \\ 
{\sl Elvang-Emparan-Mateos-Reall}~\cite{Elvang3} &\ yes${}^\dagger $ & no & {yes} & {yes}${}^\dagger $ & {yes}${}^\dagger $ & {yes}${}^\dagger $ \\
{\sl Ishihara-Matsuno}~\cite{IM} &\  yes & no & no & yes & yes & no \\
{\sl Nakagawa-Ishihara-Matsuno-Tomizawa}~\cite{NIMT} {\ }&\ yes &  no & yes${}^\dagger $ & yes${}^\dagger $ & yes${}^\dagger $ & yes${}^\dagger $ \\
{\sl Tomizawa-Ishihara-Matsuno-Nakagawa}~\cite{TIMN}  &\  yes & no & {yes} & yes & yes & yes \\
{\sl Tomizawa-Yasui-Morisawa}~\cite{TYM} &\  yes & {yes} & no & yes & yes & no \\ \hline
\end{tabular}
\caption[smallcaption]{Classification of Kaluza-Klein black holes in five-dimensional minimal supergravity: 
The six charges, $M$, $J$, $Q$, $P$, $q$ and $p$ denote, respectively, their 
mass, angular momentum, Kaluza-Klein electric charge, Kaluza-Klein magnetic charge, 
electric charge and magnetic charge.  Here the charges with a dagger `` ${}\dagger{}$ " for each solution are not independent but related by a certain constraint. }
\label{list:KK}
\end{center}
\end{table}

\medskip
{\sl Hidden symmetries} that a theory possesses  
often make it possible for us to find new solutions.  
It has been known for a long time that dimensionally reduced gravity 
(and supergravity) theories typically possess a non-compact group global 
symmetry \cite{Ehlers,MM,Geroch,CFS,CJ}, which
enables us to generate a new solution by acting this group transformation 
on a known ``seed" solution. The dimensional reduction of five-dimensional 
minimal supergravity  \cite{Cremmer} to four dimensions was performed in \cite{CN} and reconsidered in \cite{MO}, where in the latter the similarity of five-dimensional 
minimal supergravity to eleven-dimensional supergravity was emphasized.
In particular, it was shown there \cite{MO} that the $D=4$ reduced theory have $SL(2,R)$ 
symmetry precisely in the same manner as $D=4$, $N=8$ supergravity (which 
is obtained by dimensional reduction of $D=11$ supergravity \cite{CJS}) exhibits 
$E_{7(+7)}$ symmetry \cite{CJ}, which is the continuous version of U-duality \cite{HT}
(see \cite{OP} for a review)
in typeII string theory.  Also, in the presence of two space-like commuting Killing vector
fields, it is described by the $G_{2(+2)}/SO(4)$ (or $/SO(2,2)=/[SL(2,R)\times SL(2,R)]$ 
if one of the Killing 
vector is time-like) sigma model couple to gravity analogous to the $E_{8(+8)}/SO(16)$ 
sigma model \cite{MarcusSchwarz} arising in the dimensional reduction of $D=11$ 
supergravity to three dimensions.
The similarity of the structures of these two sigma models is made manifest by 
the use of Freudenthal's realization of the Lie algebras, $G_2$ and $E_8$ 
\cite{MizoguchiE10,MO,Germar}.

\medskip
So far, various types of black hole solutions in  five-dimensional theories have been derived with the help of recent development of solution-generation techniques, such as non-linear sigma model 
approach~\cite{Rasheed,FGPS,Giusto-Saxena,CEFGS,GRS,BCCGSW,CBJV,TYM,GS1,GS2}, 
as well as supersymmetric black hole solutions
~\cite{Gauntlett0,BMPV,Elvang}. 
In this paper, we utilize the above-mentioned $SL(2,R)$ symmetry of the dimensionally reduced
five-dimensional minimal supergravity to four dimensions. The bosonic sector consists
of  two Maxwell fields, a massless axion and a dilaton, all  coupled to gravity.
As was shown in Ref.~\cite{MO}, the equations of motions (derived by the dimensional reduction) are invariant under the action of a global $SL(2,R)$ group, by which Maxwell's fields are related to Kaluza-Klein's electromagnetic fields.
This {\sl $SL(2,R)$-duality}  
admits us to generate a new solution in (the bosonic sector of) five-dimensional minimal supergravity by stating from a certain known solution in the same theory 
\footnote{This procedure is essentially the same as generating a U-duality multiplet; 
in this paper we use the transformation to obtain {\sl explicit} forms of Kaluza-Klein 
black hole solutions.}.
One of the advantages of this $D=4$ $SL(2,R)$ duality  is that, unlike the $D=3$ ($G_{2(+2)}$) 
duality, one need not integrate back the dualized scalar potentials to recover the $U(1)$ gauge 
fields, which is in general a complicated problem. Another advantage is that this transformation 
preserves the asymptotic behavior of the solutions at infinity. 
The similar type of  electric-magnetic duality in non-linear electromagnetism has 
been studied in earlier works~\cite{Rasheed-Gibbons1,Rasheed-Gibbons2}.

\medskip
The known examples of Kaluza-Klein black holes within (the bosonic sector of) 
 five-dimensional minimal supergravity are summarized in TABLE~\ref{list:KK} (except trivial solutions such as black strings), where 
they are classified according to their conserved charges (mass, angular momentum, Kaluza-Klein electric/magnetic charges and electric/magnetic charges of the Maxwell field) 
which the dimensionally reduced four-dimensional black holes carry.
As seen in the list, the most general black hole solutions having six independent charges, which are 
expected to exist, have not been discovered so far. 
\footnote{Besides the list, there are many direct-product black string solutions 
known to exist, some of which have five independent charges \cite{CdBSV}.}
The purpose of this paper is to develop a new solution-generation 
technique by applying the framework of the $SL(2,R)$-duality to a known 
starting-point solution in five-dimensional theories, in order to find exact solutions of such general black hole solutions with six charges 
in five-dimensional minimal supergravity. As we will see, we will indeed find a six-charge solution starting from a four-charge (the Rasheed) 
solution, but unfortunately these six charges are not independent but related by a 
single constraint. We expect, however, that by further applying another different 
$SL(2,R)$ transformation associated with a different choice of a Killing vector 
we may be able to find full six-parameter solutions; the work is in progress.

\medskip
The remainder of this paper is organized as follows: 
In the next section, we will briefly describe our strategy for the solution-generation 
technique and write down some necessary formulas, such as the equations of motion and 
the definitions of relevant fields. 
In Section III, by acting 
the $SL(2,R)$ transformation 
on a certain seed solution, we write down some necessary formulas.
In Section IV, we provide some necessary information concerning the Rasheed solution, which we use as a seed in this paper. 
In Section V, we present new black hole solutions and 
study some basic properties, such as asymptotic charges, its topology and regularity.
Section VI is devoted to summarizing our results and discussing our formalism.

\medskip
Notational remark: In this paper we use many ``$A$"'s for different quantities; to avoid 
confusion we summarize their definitions here: $A_M$ denotes the five-dimensional 
Maxwell field in minimal supergravity. $A_\mu$ is the four-dimensional spacetime component, 
and $A_5$ is the fifth component and becomes an axion upon reduction to four dimensions. 
$A'_\mu$ is the Kaluza-Klein-independent gauge potential, and 
$\tilde A_\mu$ is the dual gauge potential. All these are transformed by the
$SL(2,R)$ transformation and have corresponding ``new" quantities, which are 
indicated by the superscript ${}^{new}$.
On the other hand, by just ``$A$" we mean the scalar function used in the metric of the Rasheed 
solution.  This $A$ does not change under the $SL(2,R)$ transformation, and hence 
there is no ``new" quantity for {\sl this} $A$ (see below). 
The symbol $2A^{(Rasheed)}_\mu$, which only appears in the remark 
above (\ref{H_munu}) and a footnote, is replaced in this paper with $B_\mu$, 
the Kaluza-Klein gauge potential. Finally, the calligraphic ${\cal A}$ 
is used  as the five-dimensional 1-form $A_M dx^M$, and 
$A^{new}\equiv A^{new}_\mu dx^\mu$ as well as ${A'}^{new}\equiv {A'}^{new}_\mu dx^\mu$
are also 1-forms. (Without ``new" we do not use such an abbreviation, otherwise it would 
conflict with Rasheed's scalar function $A$.)

\section{D=4 SL(2,R) duality}
In this section we review the $SL(2,R)$ duality symmetry \cite{MO} 
of five-dimensional minimal supergravity \cite{Cremmer,CN}
dimensionally reduced to four dimensions. The bosonic Lagrangian that we use is
\beqa
{\cal L}
=E^{(5)}\left(R^{(5)}-\frac14 F_{MN}F^{MN}\right)
-\frac1{12\sqrt{3}}\epsilon^{MNPQR}F_{MN}F_{PQ}A_R.
\label{5DL}
\eeqa
The indices $M,N,\ldots$ run over $0,1,2,3$ and $5$.
$F_{MN}\equiv\partial_{M}A_{N}-\partial_{N}A_{M}$. 
$E^{(5)}$ is the determinant of the vielbein $E^{(5)A}_{~M}$, related 
to  the five-dimensional metric $G^{(5)}_{MN}$ as
\beqa
G^{(5)}_{MN}&=&E^{(5)A}_{~M}E^{(5)B}_{~N}\eta_{AB},\\
\eta_{AB}&\equiv&\mbox{diag}(-1,+1,+1,+1,+1),
\eeqa
and hence $E^{(5)}=\sqrt{-G^{(5)}}$. Finally, $\epsilon^{MNPQR}$ is densitized,
taking values $\pm1$.

\medskip
We consider solutions which allow a Killing vector field, and take a suitable 
coordinate system so that the Killing vector is written as $\partial/\partial x^5$.  
Denoting the rest of the coordinates as $x^\mu$ ($\mu=0,1,2,3$), we decompose the 
vielbein and gauge field in the form
\beqa
E^{(5)A}_{~M}&=&\left(
\begin{array}{cc}
\rho^{-\frac12} E^{(4)\alpha}_{~~\mu}
&B_\mu \rho \\
0& \rho
\end{array}
\right),\\
A_M&=&(A_\mu, A_5).
\eeqa
Such a class of solutions 
satisfy equations of motion derived 
from the Lagrangian with the $x^5$ dependence dropped \cite{CN}:
\beqa
{\cal L}
&=& E^{(4)}\left(R^{(4)}
-\frac32 \partial_{{\mu}}\ln\rho \partial^{{\mu}}\ln\rho
-\frac12\rho^{-2} \partial_{{\mu}}A_5 \partial^{{\mu}}A_5
-\frac14\rho^3 B_{{\mu}{\nu}} B^{{\mu}{\nu}}
\right.\n
&&\left.
-\frac14\rho F^{(4)}_{{\mu}{\nu}}
F^{(4){\mu}{\nu}}
-\frac1{4\sqrt{3}}E^{(4)-1}
\epsilon^{{\mu}{\nu}{\rho}{\sigma}}
F_{{\mu}{\nu}}F_{{\rho}{\sigma}}A_5
\right),
\label{4DL}
\eeqa
where
\beqa
F^{(4)}_{{\mu}{\nu}}
 &\equiv& F'_{{\mu}{\nu}} + B_{{\mu}{\nu}}A_5,\\
F'_{{\mu}{\nu}}
&\equiv&\partial_{{\mu}}A'_{{\nu}}-\partial_{{\nu}}A'_{{\mu}},\\
A'_{{\mu}}&\equiv&A_{{\mu}}-B_{{\mu}} A_5,\label{eq:A'}
\eeqa
and
\beqa
B_{{\mu}{\nu}} &\equiv&  \partial_{{{\mu}}} B_{{\nu}}- \partial_{{{\nu}}} B_{{\mu}}.
\eeqa

\medskip
The scalar part of the Lagrangian (\ref{4DL}) is already in the form of 
the $SL(2,R)/U(1)$ nonlinear sigma model. To show that the whole system has an $SL(2,R)$ symmetry, 
one needs to express the degrees of freedom of the four-dimensional gauge field 
$A_\mu$ in terms of its electro-magnetic dual $\tilde A_\mu$ \cite{CJ,MO}.
To do this we add to (\ref{4DL}) a Lagrange multiplier term 
\beqa
{\cal L}_{\rm Lag.mult.}=\frac12 
\epsilon^{{\mu}{\nu}{\rho}{\sigma}}
\tilde{A}_{{\sigma}}\partial_{{\rho}}F'_{{\mu}{\nu}},
\label{LLm}
\eeqa
partially integrate it and complete the square.
Discarding the perfect square, one finds \cite{MO}
\beqa
{\cal L}+{\cal L}_{\rm Lag.mult.}
&=&E^{(4)}R^{(4)} + {\cal L}_S +{\cal L}_V,\n
{\cal L}_S
&\equiv&-E^{(4)}\left(\frac32\partial_{{\mu}}\ln \rho
                        \partial^{{\mu}}\ln \rho
                 +\frac12\rho^{-2}\partial_{{\mu}}A_5
                        \partial^{{\mu}}A_5\right),\n
{\cal L}_V
&\equiv&-\frac14E^{(4)}{\cal G}^T_{{\mu}{\nu}}
N^{{\mu}{\nu}{\rho}{\sigma}}
{\cal G}_{{\rho}{\sigma}},
\label{4d}
\eeqa
where 
\beqa
{\cal G}_{{\mu}{\nu}}
&\equiv&\left(\begin{array}{c}\tilde{A}_{{\mu}{\nu}}\\
B_{{\mu}{\nu}}\end{array} \right),\\ 
{\tilde A}_{{\mu}{\nu}}&\equiv& \partial_{{{\mu}}}{\tilde A}_{{\nu}}-\partial_{{{\nu}}}{\tilde A}_{{\mu}}.
\eeqa 
$N^{{\mu}{\nu}{\rho}{\sigma}}$ is a two-by-two matrix 
in the form
\beqa
N^{{\mu}{\nu}{\rho}{\sigma}}
&=& m~{1}^{{\mu}{\nu}{\rho}{\sigma}}
+a~(\ast)^{{\mu}{\nu}{\rho}{\sigma}},\n
V^{-1}mV^{-1}&=&K-\frac12(\Phi\Phi^*K+K\Phi^*\Phi)
+\frac14\Phi\Phi^{*2}K\Phi,\n
V^{-1}aV^{-1}&=&-\Phi^*K-\Phi+\frac12(\Phi\Phi^{*2}K+K\Phi^{*2}\Phi)
 +\frac13\Phi\Phi^*\Phi-\frac14\Phi\Phi^{*3}K\Phi,
\eeqa
where 
\beqa
{1}^{{\mu}{\nu}{\rho}{\sigma}}
&\equiv& \frac12\left(G^{(4){\mu}{\rho}}G^{(4){\nu}{\sigma}}
-G^{(4){\nu}{\rho}}G^{(4){\mu}{\sigma}}\right),\\ 
(\ast)^{{\mu}{\nu}{\rho}{\sigma}}
&\equiv&\frac12 E^{(4)-1}\epsilon^{{\mu}{\nu}{\rho}{\sigma}},
\eeqa
with
\beqa
V \equiv \left(\begin{array}{cc}\rho^{-\frac12}& 0 \\
 0 &\rho^{\frac32}\end{array}
\right),~~
 \Phi\equiv\left(\begin{array}{cc}
 0 &\sqrt{3}\phi\\
 \sqrt{3}\phi& 0 \end{array}
\right),~~
\Phi^*\equiv\left(\begin{array}{cc} 2\phi& 0 \\ 0 &0\end{array}
\right),
\eeqa
\beqa
K \equiv(1+\Phi^{*2})^{-1},~ 
\phi\equiv\frac1{\sqrt{3}}\rho^{-1}A_5.
\eeqa
The square term discarded in (\ref{4d}) gives a relation between 
$A_\mu$ and $\tilde A_\mu$:
\beqa
\ast F^{(4)}_{\mu\nu}&=&-M_{\mu\nu\mu'\nu'} \rho^{-1}
\left(\frac1{\sqrt 3} A_5^2 B^{\mu'\nu'}-\tilde{A}^{\mu'\nu'}\right),\\
(M^{-1})^{\mu\nu\mu'\nu'}&\equiv&\left(
1+\frac2{\sqrt 3}\rho^{-1}A_5 (\ast)
\right)^{\mu\nu\mu'\nu'},
\eeqa
or
\beqa
\tilde{A}_{\mu\nu}&=&\rho(\ast F^{(4)})_{\mu\nu}-\frac 2{\sqrt 3}A_5 F^{(4)}_{\mu\nu}
+\frac1{\sqrt 3}A_5^2 B_{\mu\nu}.
\eeqa

The dimensionally reduced system has two independent four-dimensional 
gauge fields, the Kaluza-Klein gauge field $B_\mu$ ($=2 A_\mu^{(Rasheed)}$ 
in the notation of \cite{Rasheed}) and the dual of the four-dimensional component of 
the gauge field $\tilde A_\mu$. The $SL(2,R)$ symmetry mixes these two {\em and} their 
electromagnetic duals together into a linear combination. More precisely, 
if one defines 
\beqa
{\cal H}_{\mu\nu}~\equiv~\left(
\begin{array}{c}
{\cal H}_{\mu\nu}^{\tilde{A}}\\
{\cal H}_{\mu\nu}^B
\end{array}
\right)
&\equiv&m(\ast {\cal G})_{\mu\nu} -a~{\cal G}_{\mu\nu}
\label{H_munu}
\eeqa
and write the four ``field strengths" in a single column vector 
\beqa
{\cal F}_{\mu\nu}&\equiv&\left(
\begin{array}{c}
{\cal G}_{\mu\nu}\\
{\cal H}_{\mu\nu}
\end{array}
\right),
\eeqa
then it can be shown that ${\cal F}_{\mu\nu}$ satisfies
\beqa
{\cal F}_{\mu\nu}&=&
\Omega{\cal V}_+^T {\cal V}_-^T\cdot {\cal V}_- {\cal V}_+(\ast{\cal F})_{\mu\nu},
\eeqa
where
\beqa
{\cal V}_+=\left(
\begin{array}{cc}
~V~~&\\
&V^{-1}
\end{array}
\right),
\quad{\cal V}_-=\exp\left(
\begin{array}{cc}
&-\Phi^*\\
-\Phi&
\end{array}
\right),
\eeqa

\noindent
and
\beqa
\Omega&\equiv&
\left(
\begin{array}{cccc}
&&-1~&~\\
&&~&-1\\
~1~&~~&&\\
~~&~1~&&
\end{array}
\right)
\eeqa
is the invariant matrix of $Sp(4)$.

\medskip
To see the $SL(2,R)$ invariance in the vector sector, it is convenient to 
introduce representation matrices of the $SL(2,R)$ Chevalley generators 
embedded in $Sp(4)$:
\beqa
E\equiv\left(
\begin{array}{cccc}
~0~&\sqrt{3}&&\\
&~0~&2&\\
&&~0~&\sqrt{3}~\\
&&&~0~
\end{array}
\right),~
F\equiv\left(
\begin{array}{cccc}
~0~&&&\\
\sqrt{3}&~0~&&\\
&2&~0~&\\
&&\sqrt{3}&~0~
\end{array}
\right),~
H\equiv\left(
\begin{array}{cccc}
~3~&~~&~~&~~\\
~~&~1~&~~&~~\\
~~&~~&-1&~~\\
~~&~~&~~&-3
\end{array}
\right),
\eeqa
and their similarity transformations
\beqa
E'\equiv P^{-1}EP,~~
F'\equiv P^{-1}FP,~~
H'\equiv P^{-1}HP,
\label{4rep}
\eeqa
\beqa
P~=~ (P^{-1})^T&\equiv&
\left(
\begin{array}{cccc}
~0~&~0~&~0~&~1~\\
~1~&~0~&~0~&~0~\\
~0~&~0~&~1~&~0~\\
~0~&~1~&~0~&~0~
\end{array}
\right).
\eeqa
Then we can write
\beqa
{\cal V}_- {\cal V}_+ &=& \exp(\phi E') \exp(-\log \rho^{\frac12} H').
\eeqa
Let $\Lambda\in SL(2,R)$
generated by $E'$, $F'$ and $H'$. Then
\beqa
\Lambda^{-1}{\cal F}_{\mu\nu}&=&\Lambda^{-1}\Omega{\cal V}_+^T{\cal V}_-^T{\cal V}_-{\cal V}_+ (\ast{\cal F})_{\mu\nu}\n
&=&\Omega(\Lambda^{T}{\cal V}_+^T{\cal V}_-^T{\cal V}_-{\cal V}_+ \Lambda) \Lambda^{-1} (\ast{\cal F})_{\mu\nu}.
\eeqa
 Therefore the equations of motion and the Bianchi identity are invariant under
\beqa
{\cal F}_{\mu\nu}&\mapsto&\Lambda^{-1}{\cal F}_{\mu\nu},\label{eq:Ftr}\\
{\cal V}_-{\cal V}_+ &\mapsto& {\cal V}_-{\cal V}_+ \Lambda. \label{VVtoVVU-1}
\eeqa
Also using  ${\cal R}=({\cal V}_-{\cal V}_+)^T{\cal V}_-{\cal V}_+$, the scalar Lagrangian can be written 
as
\beqa
{\cal L}_S
=\frac3{40} E^{(4)} {\rm Tr}\partial_{{\mu}}{\cal R}^{-1}
\partial^{{\mu}}{\cal R},
\label{L_S}
\eeqa
which is also invariant under (\ref{VVtoVVU-1}). The trace in (\ref{L_S}) is taken in the 
{\bf 4} representation ({\ref{4rep}}).

\medskip
This is precisely analogous to the dimensional reduction of 
eleven-dimensional supergravity \cite{CJS} to four-dimensions \cite{CJ},
where the coset is then $E_{7(+7)}/SU(8)$ and twenty-eight vector fields are 
similarly doubled and transform as {\bf 56} of $E_7$ 
represented as $Sp(56)$ matrices \cite{HT}.
The four gauge fields ($\tilde A_\mu$, $B_\mu$ and their electro-magnetic duals) 
transform as {\bf 4} (that is, spin-$\frac32$) representation of $SL(2.R)$. Therefore,
$\tilde A_\mu$, $B_\mu$ are not on equal footing; upon further dimensional reduction 
to three dimensions, $\tilde A_\mu$($B_\mu$) corresponds to a short(long) root 
of the non-simply-laced Lie algebra $G_2$ \cite{Germar}.

\medskip
Finally, since ${\cal H}_{\mu\nu}^{\tilde{A}}$ (\ref{H_munu}) 
 is the ``dual of dual",
it is essentially the field strength of the original gauge field before the dual is taken.
More precisely, it can be shown that
\beqa
{\cal H}_{\mu\nu}^{\tilde{A}}&=&-F'_{{\mu}{\nu}}.
\eeqa

\section{Solution Generation}
Utilizing the $SL(2,R)$ symmetry which the field equations in the dimensionally reduced four-dimensional spacetime have, 
one can generate new solutions from already known solutions within five-dimensional minimal supergravity; 
the latter 
are often called {\sl seed solutions} in the context of solution-generation techniques. 
The strategy to generate a new solution consists in the following steps: 
First, to reduce the five-dimensional spacetime to four dimensions, one must choose a Killing vector $\xi=\partial/\partial w$ so that the metric and gauge potential 1-form does not depends on the parameter $w$  
(in this paper, we take it to be the $U(1)$ generator of compactified fifth direction $\partial/\partial x^5$).
The dimensional reduction of the five-dimensional metric and the Maxwell field yields the four-dimensional Einstein theory with
two Maxwell fields, a dilaton, and an axion. From the dimensionally reduced metric and the gauge fields, one  reads off 
the dilaton, axion $(\rho,A_5)$ and the Maxwell field, Kaluza-Klein gauge field $(A_\mu,B_\mu)$. 
In terms of the scalar fields and vector fields for a seed solution, one 
constructs the matrix ${\cal R}$ and the field strength vector ${\cal F}_{\mu\nu}$.
Next,  applying a suitable $SL(2,R)$-transformation to the matrix ${\cal R}$ and the four-vector ${\cal F}_{\mu\nu}$ 
representing the seed solution, i.e. ${\cal R}\to \Lambda^T{\cal R}\Lambda$, ${\cal F}_{\mu\nu}\to \Lambda^{-1}{\cal F}_{\mu\nu}$ for $\Lambda\in SL(2,R)$, one can get 
a new matrix ${\cal R}^{new}$ and a four-vector ${\cal F}^{new}_{\mu\nu}$. 
It is straightforward to read off the dilaton and axion for the new solution from the transformed matrix ${\cal R}^{new}$. 
Also, from the transformed field strength vector ${\cal F}^{new}_{\mu\nu}$, one can read off the field strengths $(B^{new}_{\mu\nu},{F'}^{new}_{\mu\nu})$ of the transformed solution.  
Integrating $B^{new}_{\mu\nu}$, one gets the Kaluza-Klein gauge field $B^{new}_\mu$. Further, integrating ${F'}^{new}_{\mu\nu}$ and combining Eq.~(\ref{eq:A'}), one obtains the Maxwell field $A^{new}_{\mu}$ of 
the transformed solution.

\subsection{Scalar fields -axion and dilaton-}
To derive the transformation rules for the two scalar fields, it is more convenient to 
use the $2\times2$-matrix representation of $SL(2,R)$ rather than the $4\times4$-matrix representation 
introduced in the previous section. The isomorphism $\pi$ maps the generators 
as follows:
\beqa
\pi(E')=\left(
\begin{array}{cc}
&1\\0&
\end{array}
\right),~~
\pi(F')=\left(
\begin{array}{cc}
&0\\1&
\end{array}
\right),~~
\pi(H')=\left(
\begin{array}{cc}
1&\\&-1
\end{array}
\right).
\label{isomorphism}
\eeqa
In general, any group element of $SL(2,R)$ can be decomposed in this defining representation as
\beqa
\left(
\begin{array}{cc}
a&b\\c&d
\end{array}
\right)&=&\left(
\begin{array}{cc}
1&-\alpha\\0&1
\end{array}
\right)
\left(
\begin{array}{cc}
\delta&0\\0&\delta^{-1}
\end{array}
\right)
\left(
\begin{array}{cc}
1&0\\ -\beta&1
\end{array}
\right)
\eeqa
if $d\neq 0$, where $a$, $b$, $c$, $d$, $\alpha$, $\beta$ and $\delta$ are all real numbers 
with $ad-bc=0$ and $\delta\neq 0$. Also, the element with $d=0$ can be obtained by blowing up 
the singularity that occurs in the limit $\delta\rightarrow 0$. 
Therefore, a general choice for $\Lambda$ is
\beqa
\Lambda=e^{-\alpha E'}e^{(\log\delta) H'}e^{-\beta F'}.
\eeqa
However, it turns out that the middle factor coming from the Cartan subalgebra does not 
produce any new solution but only changes the values of the parameters of the solution.
Therefore 
we choose
\beqa
\Lambda=e^{-\alpha E'}e^{-\beta F'},
\eeqa
and compute ${\cal R}^{new}$,
following Eq. (\ref{VVtoVVU-1}). 
Using the isomorphism (\ref{isomorphism}), we find
\begin{eqnarray}
\pi({\cal R})&=&
\left(
\begin{array}{cc}
\rho^{-1}& -\frac{1}{\sqrt{3}}\rho^{-1}A_5  \\
-\frac{1}{\sqrt{3}}\rho^{-1}A_5 & \frac{1}{3}\rho^{-1}A_5^2+\rho
\end{array}
\right)\\
\to \pi({\cal R}^{new})=\pi(\Lambda^T {\cal R}\Lambda) 
&=&\left(
\begin{array}{cc}
\left(1+\beta(\alpha+\frac{ A_5}{\sqrt{3}})\right)^2\rho^{-1}+\beta^2\rho & -\left(1+\beta(\alpha+\frac{A_5}{\sqrt{3}})\right)\left(\alpha+\frac{A_5}{\sqrt{3}}\right)\rho^{-1}-\beta\rho\\
-\left(1+\beta(\alpha+\frac{A_5}{\sqrt{3}})\right)\left(\alpha+\frac{A_5}{\sqrt{3}}\right)\rho^{-1}-\beta\rho & (\alpha+\frac{A_5}{\sqrt{3}})^2 \rho^{-1}+\rho\label{eq:trR}
\end{array}
\right).\n
\end{eqnarray}

In particular, if we transform vacuum solutions of five-dimensional Einstein theory,
we can set the axion $A_5$ to be $0$. 
In this case, from Eq.~(\ref{eq:trR}), the transformed dilaton field and axion field read:
\begin{eqnarray}
\rho^{new}=\frac{\rho}{(1+\alpha\beta)^2+\beta^2\rho^2},\label{eq:newrho}
\end{eqnarray}
\begin{eqnarray}
A_5^{new}=\sqrt{3}\frac{\alpha(1+\alpha\beta)+\beta\rho^2}{(1+\alpha\beta)^2+\beta^2\rho^2}.\label{eq:newA5}
\end{eqnarray}

\subsection{Vector fields - Kaluza-Klein gauge field and Maxwell field -}
Next, we consider the sector of two vector fields, the 
Kaluza-Klein gauge field and the Maxwell field.
The transformation (\ref{eq:Ftr}) with $\Lambda=e^{-\alpha E'}e^{-\beta F'}$
yields
\begin{eqnarray}
&&{\cal F}_{\mu\nu}=
\left(
\begin{array}{c}
\tilde A_{\mu\nu}\\
B_{\mu\nu}\\
-F'_{\mu\nu}\\
{\cal H}_{\mu\nu}^B
\end{array}
\right)\\
&&\to {\cal F}^{new}_{\mu\nu}=
\Lambda^{-1}{\cal F}_{\mu\nu}
=\left(
\begin{array}{cccc}
 1+3 \alpha  \beta  & \sqrt{3} \alpha ^2 (1+\alpha  \beta ) & \alpha  (2+3 \alpha  \beta ) & \sqrt{3} \beta  \\
 \sqrt{3} \beta ^2 (1+\alpha  \beta ) & (1+\alpha  \beta )^3 & \sqrt{3} \beta  (1+\alpha  \beta )^2 & \beta ^3 \\
 \beta  (2+3 \alpha  \beta ) & \sqrt{3} \alpha  (1+\alpha  \beta )^2 & 1+4 \alpha  \beta +3 \alpha ^2 \beta ^2 & \sqrt{3} \beta ^2 \\
 \sqrt{3} \alpha  & \alpha ^3 & \sqrt{3} \alpha ^2 & 1
\end{array}
\right)
\left(
\begin{array}{c}
\tilde A_{\mu\nu}\\
B_{\mu\nu}\\
-F'_{\mu\nu}\\
{\cal H}_{\mu\nu}^B
\end{array}
\right).
\end{eqnarray}
Again, if the seed solution has no Maxwell fields,  the field-strength vector ${\cal F}_{\mu\nu}$ 
is simplified to
\beqa
{\cal F}_{\mu\nu}&=&
\left(
\begin{array}{c}
0\\
B_{\mu\nu}\\
0\\
\rho^3(*B)_{\mu\nu}
\end{array}
\right),
\eeqa
and the new Kaluza-Klein gauge field strength 
can be written in the form:
\begin{eqnarray}
B^{new}_{\mu\nu}=(1+\alpha\beta)^3B_{\mu\nu}+\beta^3\rho^3(*B)_{\mu\nu}.\label{eq:Bnew0}
\end{eqnarray}
Here, let us define the dual $1$-form $\tilde B=\tilde B_\mu dx^\mu$ for the $1$-form $B=B_\mu dx^\mu$ 
\footnote{This notation ``1-form $B=B_\mu dx^\mu$" is only used in this subsection. 
To avoid confusion with Rasheed's $B$, we will henceforth write 
it as $B_\mu dx^\mu$ explicitly.}
by
\begin{eqnarray}
d\tilde B=\rho^3*dB.\label{eq:tildeB}
\end{eqnarray}
The equation of motion for the vector $B_\mu$ assures existence of such a $1$-form $\tilde B$. 
Then, for the transformed solutions, in terms of $\tilde B$, the Kaluza-Klein gauge potential $1$-form 
$B^{new}=B^{new}_\mu dx^\mu$  can be written as
\begin{eqnarray}
B^{new}=(1+\alpha\beta)^3B+\beta^3\tilde B.\label{eq:Bnew1}
\end{eqnarray}

On the other hand, from the third column of the four-vector ${\cal F}^{new}_{\mu\nu}$, 
it turns out that 
\begin{eqnarray}
-F_{\mu\nu}^{\prime new}=\sqrt{3}\alpha(1+\alpha\beta)^2B_{\mu\nu}+\sqrt{3}\beta^2 \rho^3(*B)_{\mu\nu},
\end{eqnarray}
which can be integrated to give
\begin{eqnarray}
A^{\prime new}=-\sqrt{3}\alpha(1+\alpha\beta)^2B-\sqrt{3}\beta^2 \tilde B.
\end{eqnarray}
Recall (Eq. (\ref{eq:A'})) that the $1$-form $A^{\prime new}$  
can be written as
\begin{eqnarray}
A^{\prime new}=A^{new}-A_5^{new}B^{new}.
\end{eqnarray}
Therefore, the new
gauge potential $1$-form $A^{new}=A^{new}_\mu dx^\mu$ 
is
\begin{eqnarray}
A^{new}=\left[(1+\alpha\beta)^3A_5^{new}-\sqrt{3}\alpha(1+\alpha\beta)^2\right]B+\left[\beta^3A_5-\sqrt{3}\beta^2\right]\tilde B.\label{eq:Anew1}
\end{eqnarray}

\section{Rasheed solution}
In the following section, using the $SL(2,R)$-transformation mentioned in the previous section,
we will generate a new rotating black hole solution, starting from the Rasheed solution~\cite{Rasheed}.
Hence, in this section, we briefly review the Rasheed solutions in five-dimensional pure gravity.
The metric of the Rasheed solution is given by
\begin{eqnarray}
ds^2&=&\frac{B}{A}(dx^5+
B_\mu dx^\mu)^2+\sqrt{\frac{A}{B}}ds^2_{(4)},\label{eq:Rasheed}
\end{eqnarray}
where the four-dimensional (dimensionally reduced) metric is given by
\begin{eqnarray}
ds^2_{(4)}=-\frac{f^2}{\sqrt{AB}}(dt+\omega^0{}_\phi d\phi)^2+\frac{\sqrt{AB}}{\Delta}dr^2+\sqrt{AB}d\theta^2+\frac{\sqrt{AB}\Delta}{f^2}\sin^2\theta d\phi^2.\label{eq:4metric}
\end{eqnarray}
Here the functions $(A,B,C,\omega^0{}_\phi,\omega^5{}_{\phi},f^2,\Delta)$ and the $1$-form 
$B_{\mu}$ 
\footnote{To avoid confusion with the Maxwell field, we denote in this paper 
the Kaluza-Klein vector field as $B_{\mu}$,  instead of $2 A_\mu^{(Rasheed)}$ 
used in the original article \cite{Rasheed}.}
are
\begin{eqnarray}
&&A=\left(r-\frac{\Sigma}{\sqrt{3}}\right)^2-\frac{2P^2\Sigma}{\Sigma-\sqrt{3}M}+a^2\cos^2\theta+\frac{2JPQ\cos\theta}{(M+\Sigma/\sqrt{3})^2-Q^2},\\
&&B=\left(r+\frac{\Sigma}{\sqrt{3}}\right)^2-\frac{2Q^2\Sigma}{\Sigma+\sqrt{3}M}+a^2\cos^2\theta-\frac{2JPQ\cos\theta}{(M-\Sigma/\sqrt{3})^2-P^2},\\
&&C=2Q(r-\Sigma/\sqrt{3})-\frac{2PJ\cos\theta(M+\Sigma/\sqrt{3})}{(M-\Sigma/\sqrt{3})^2-P^2},\\
&&\omega^0{}_\phi=\frac{2J\sin^2\theta}{f^2}\left[r-M+\frac{(M^2+\Sigma^2-P^2-Q^2)(M+\Sigma/\sqrt{3})}{(M+\Sigma/\sqrt{3})^2-Q^2}\right],\\
&&\omega^5{}_\phi=\frac{2P\Delta \cos\theta}{f^2}-\frac{2QJ\sin^2\theta[r(M-\Sigma/\sqrt{3})+M\Sigma/\sqrt{3}+\Sigma^2-P^2-Q^2]}{f^2[(M+\Sigma/\sqrt{3})^2-Q^2]},\\
&&\Delta=r^2-2Mr+P^2+Q^2-\Sigma^2+a^2,\\
&&f^2=r^2-2Mr +P^2+Q^2-\Sigma^2+a^2\cos^2\theta,\\
&& 
B_\mu dx^\mu=\frac{C}{B}dt+\left(\omega^5{}_\phi+\frac{C}{B}\omega^0{}_\phi\right)d\phi,\label{eq:RasheedB}
\end{eqnarray}
where $B_\mu$ describes the electromagnetic vector potential derived by dimensional reduction to four dimension. Here the constants, $(M,P,Q,J,\Sigma)$, mean the mass, Kaluza-Klein magnetic charge, Kaluza-Klein electric charge, angular momentum along four dimension and dilaton charge, respectively, which are parameterized by the two parameters $(\hat\alpha,\hat\beta)$
\begin{eqnarray}
&&M=\frac{(1+\cosh^2\hat\alpha\cosh^2\hat\beta)\cosh\hat\alpha}{2\sqrt{1+\sinh^2\hat\alpha\cosh^2\hat\beta}}M_k,\\
&&\Sigma=\frac{\sqrt{3}\cosh\hat\alpha(1-\cosh^2\hat\beta+\sinh^2\hat\alpha\cosh^2\hat\beta)}{2\sqrt{1+\sinh^2\hat\alpha\cosh^2\hat\beta}}M_k,\\
&&Q=\sinh\hat\alpha\sqrt{1+\sinh^2\hat\alpha\cosh^2\hat\beta}\ M_k,\\
&&P=\frac{\sinh\hat\beta \cosh\hat\beta}{\sqrt{1+\sinh^2\hat\alpha\cosh^2\hat\beta}}M_k,\\
&&J= \cosh\hat\beta \sqrt{1+\sinh^2\hat\alpha\cosh^2\hat\beta\ }aM_k.
\end{eqnarray}
Note that all the above parameters are not independent since they are related through the equation
\begin{eqnarray}
\frac{Q^2}{\Sigma+\sqrt{3}M}+\frac{P^2}{\Sigma-\sqrt{3}M}=\frac{2\Sigma}{3},\label{eq:constraint}
\end{eqnarray}
and the constant $M_k$ is written in terms of these parameters
\begin{eqnarray}
M_k^2=M^2+\Sigma^2-P^2-Q^2.
\end{eqnarray}
The constant $J$ is also related to $a$ by
\begin{eqnarray}
J^2=a^2\frac{[(M+\Sigma/\sqrt{3})^2-Q^2][(M-\Sigma/\sqrt{3})^2-P^2]}{M^2+\Sigma^2-P^2-Q^2}.
\end{eqnarray}

\medskip
The dilaton and axion fields for the Rasheed solution are, respectively,
\begin{eqnarray}
\rho=\sqrt{\frac{B}{A}},\quad A_5=0.\label{eq:rhonew}
\end{eqnarray}
The Kaluza-Klein gauge field and Maxwell field are, respectively,
\begin{eqnarray}
B_\mu dx^\mu=\frac{C}{B}dt+\left(\omega^5{}_\phi+\frac{C}{B}\omega^0{}_\phi\right)d\phi,\quad 
A_\mu dx^\mu=0.
\end{eqnarray}

\section{New solutions}
Performing a series of the $SL(2,R)$ transformations $\Lambda=e^{-\alpha E'}e^{-\beta F'}$ ($\alpha,\beta$: constants) for the Rasheed solutions in five-dimensional Einstein theory, we can obtain new solutions in the bosonic sector of five-dimensional minimal supergravity.
From Eqs.~(\ref{eq:rhonew}), (\ref{eq:newrho}) and (\ref{eq:newA5}), the dilaton field and axion field for new solutions are given by, respectively,
\begin{eqnarray}
\rho^{new}=\frac{\sqrt{B/A}}{(1+\alpha\beta)^2+ \beta^2B/A},
\end{eqnarray}
\begin{eqnarray}
A_5^{new}=\frac{\sqrt{3}(\alpha+\alpha^2\beta+\beta B/A)}{(1+\alpha\beta)^2+\beta^2B/A}.
\end{eqnarray}
Integrating Eq.~(\ref{eq:tildeB}) for the Rasheed solutions (see Eq.~(\ref{eq:RasheedB})) and further using the formulas~(\ref{eq:Bnew1}) and (\ref{eq:Anew1}), we see that the Kaluza-Klein $U(1)$ field and Maxwell $U(1)$ field for the new solutions are written as
\begin{eqnarray}
B^{new}_\mu=(1+\alpha\beta)^3B_\mu+\beta^3 \tilde B_\mu,
\end{eqnarray}
\begin{eqnarray}
A^{new}_\mu=(1+\alpha\beta)^2\left[(1+\alpha\beta)A_5^{new}-\sqrt{3}\alpha\right]B_\mu+\beta^2(\beta A_5^{new}-\sqrt{3})\tilde B_\mu,
\end{eqnarray}
where
\begin{eqnarray}
B_\mu dx^\mu=\frac{C}{B}dt+\left(\omega^5{}_\phi+\frac{C}{B}\omega^0{}_\phi\right)d\phi,
\end{eqnarray}
\begin{eqnarray}
\tilde B_\mu dx^\mu&=&\frac{2P(r+\Sigma/\sqrt{3})+2JQ \frac{(M-\Sigma/\sqrt{3})}{(M+\Sigma/\sqrt{3})^2-Q^2}\cos\theta}{A}dt \nonumber\\
&-&\left(2Q\cos\theta+\frac{2JP\frac{(M+\Sigma/\sqrt{3})}{(M-\Sigma/\sqrt{3})^2-P^2}r+2JP \frac{P^2-Q^2+\Sigma(\sqrt{3}M-\Sigma/3)}{(M-\Sigma/\sqrt{3})^2-P^2}+2a^2Q \cos\theta}{A}\sin^2\theta\right) d\phi.
\end{eqnarray}
To summarize, in terms of the two parameters $(\alpha,\beta)$, the metric is in the form:
\begin{eqnarray}
ds^2=\frac{AB}{(A(1+\alpha\beta)^2+\beta^2 B)^2}\left[dx^5+(1+\alpha\beta)^3B_\mu dx^\mu+\beta^3\tilde B_\mu dx^\mu\right]^2+\frac{A(1+\alpha\beta)^2+\beta^2 B}{\sqrt{AB}}ds^2_{(4)},
\end{eqnarray}
where the four-dimensional metric is the same as that of the seed solutions (given by Eq.~(\ref{eq:4metric})).

\medskip
In order that the metric takes the {\sl standard} asymptotic form, let us now consider the rescale of the coordinates:
\begin{eqnarray}
N^{\frac{1}{2}} dt\to dt,\ \frac{dx^5}{N}\to dx^5,\ N^{\frac{1}{2}} dr\to dr,
\end{eqnarray}
and the redefinition of the parameters:
\begin{eqnarray}
N^{\frac{1}{2}} M\to M,\ N J\to J,\ N^{\frac{1}{2}} Q\to Q,\ N^{\frac{1}{2}} P\to P,\ N^{\frac{1}{2}} a\to a,\ N^{\frac{1}{2}} \Sigma\to\Sigma,
\end{eqnarray}
where  $N=\beta^2+\gamma^2$ ($\gamma\equiv 1+\alpha\beta$). In this case, we note that
\begin{eqnarray}
N A\to A,\ N B\to B,\ N C\to C,\ N^{\frac{1}{2}} \omega^0{}_\phi\to\omega^0{}_\phi,\ N^{\frac{1}{2}}\omega^5{}_\phi\to\omega^5{}_\phi,\ N \Delta\to\Delta,\ N f^2\to f^2.
\end{eqnarray}
The rescaled metric and gauge field are written in the form:
\begin{eqnarray}
ds^2=\frac{AB}{(A\tilde\gamma^2+\tilde\beta^2 B)^2}\left[dx^5+\tilde\gamma^3B_\mu dx^\mu+\tilde\beta^3\tilde B_\mu dx^\mu\right]^2+\frac{A\tilde\gamma^2+\tilde\beta^2 B}{\sqrt{AB}}ds^2_{(4)}\label{eq:newmetric}
\end{eqnarray}
and
\begin{eqnarray}
{\cal A}~\equiv~A_Mdx^M
=\left[(\tilde\gamma^3B_\mu +\tilde\beta^3\tilde B_\mu )A^{new}_5-\sqrt{3}N^{\frac{1}{2}}(\alpha\tilde\gamma^2B_\mu+\tilde\beta^2\tilde B_\mu)\right]dx^\mu+A^{new}_5dx^5,
\end{eqnarray}
where the four-dimensional metric is also invariant under the rescale (given by Eq.~(\ref{eq:4metric})).
The dilaton field and axion are also written as, respectively,
\begin{eqnarray}
\rho^{new}=\frac{\sqrt{AB}}{A\tilde\gamma^2+\tilde\beta^2 B},
\end{eqnarray}
\begin{eqnarray}
A_5^{new}=\frac{\sqrt{3}}{\tilde\beta}\left(N^{\frac{1}{2}}-\frac{\tilde\gamma}{\tilde\gamma^2+\tilde\beta^2\rho^2}\right).
\end{eqnarray}
Here, the constants $\tilde\beta$ and  $\tilde\gamma$ are defined by
\begin{eqnarray}
\tilde \gamma=N^{-\frac{1}{2}}\gamma,\ \tilde\beta=N^{-\frac{1}{2}}\beta.
\end{eqnarray}

\subsection{Asymptotic structure and conserved charges}

Next we study the asymptotics of the obtained solutions.
The metric at infinity, $r\to \infty$, behaves as
\begin{eqnarray}
ds^2&\simeq& \left(-1+\frac{-\frac{4\tilde \beta^2\Sigma}{\sqrt{3}}+2M+\frac{2\Sigma}{\sqrt{3}}}{r}+{\cal O}(r^{-2})\right)dt^2+2\left(\frac{2\tilde\beta^3P+2\tilde\gamma^3 Q}{r}+{\cal O}(r^{-2})\right)dtdx^5\nonumber\\
    & &+2\left(\frac{(2\tilde \gamma^3P-2\tilde\beta^3Q)(2\tilde\gamma^3Q+2\tilde\beta^3P)\cos\theta-J \sin^2\theta}{r}+{\cal O}(r^{-2})\right)dtd\phi\nonumber\\
    & &+\left(1+\frac{4(\tilde\gamma^2-\tilde\beta^2)\Sigma}{\sqrt{3}r}+{\cal O}(r^{-2})\right)(dx^5)^2+r^2 \sin^2\theta\left(1+{\cal O}(r^{-1})\right)d\phi^2+2\left(2\tilde\gamma^3P-2\tilde\beta^3Q+{\cal O}(r^{-1})\right)dx^5 d\phi\nonumber\\
&&+\left(1+{\cal O}(r^{-1})\right)dr^2+r^2\left(1+{\cal O}(r^{-1})\right)d\theta^2,
\end{eqnarray}
and the gauge potential behaves as
\begin{eqnarray}
{\cal A}&\simeq& \left(\frac{2\sqrt{3}\tilde\beta\tilde\gamma[\tilde\gamma Q-\tilde\beta P]}{r}+{\cal O}(r^{-2})\right)dt+\sqrt{3}\tilde \beta^{-1}\left(N^{\frac{1}{2}}-\tilde\gamma+{\cal O}(r^{-1})\right)dx^5\nonumber\\
&&+2\sqrt{3}\left(\tilde \beta \tilde\gamma(\tilde \gamma P+\tilde\beta Q)\cos\theta+{\cal O}(r^{-1}) \right)d\phi.
\end{eqnarray}
At infinity, our solutions behave as a four-dimensional flat spacetime plus a ecirclef and hence describe a Kaluza-Klein black hole.
As shown by Ref.~\cite{Tomizawa}, this coincides with the asymptotic behaviors for a compactified spacetime in the five-dimensional minimal supergravity.
Thus, we see that the $SL(2,R)$ transformation preserves the five-dimensional Kaluza-Klein asymptotic.
On the other hand, the four-dimensional metric at infinity behaves as
\begin{eqnarray}
ds^2_{(4)}&\simeq& -\left(1-\frac{2M}{r}+{\cal O}(r^{-2})\right)dt^2-\left(\frac{2J\sin^2\theta}{r}+{\cal O}(r^{-2})\right)dtd\phi+\left(1+\frac{2M}{r}+{\cal O}(r^{-2})\right)dr^2\nonumber\\
&&+r^2(1+{\cal O}(r^{-1}))(d\theta^2+\sin^2\theta d\phi^2).
\end{eqnarray}
The gauge potentials for Kaluza-Klein and Maxwell fields behaves as, respectively,
\begin{eqnarray}
B_\mu dx^\mu\simeq \left(\frac{2\tilde \beta P+2\tilde\gamma Q}{r}+{\cal O}(r^{-2})\right)dt+\left(2(\tilde\beta Q-\tilde\gamma P)\cos\theta+{\cal O}(r^{-1})\right) d\phi,
\end{eqnarray}
\begin{eqnarray}
A_\mu dx^\mu&\simeq& \left(\frac{2\sqrt{3}\tilde\beta\tilde\gamma[\tilde\gamma Q-\tilde\beta P]}{r}+{\cal O}(r^{-2})\right)dt+2\sqrt{3}\left(\tilde \beta \tilde\gamma(\tilde \gamma P+\tilde\beta Q)\cos\theta+{\cal O}(r^{-1}) \right)d\phi.
\end{eqnarray}
Here, the constants $M$ and $J$ are the asymptotic conserved mass and angular momentum for the dimensionally reduced spacetime. 
For stationary, axisymmetric spacetimes with Killing symmetries $\xi_t^\mu=(\partial/\partial t)^\mu$ and $\xi_\phi^\mu=(\partial/\partial \phi)^\mu$, the charges can be defined over the two-surface at spatial infinity $S^2_{\infty}$ as, respectively, 
\begin{eqnarray}
&&M=-\frac{1}{8\pi}\int_{S^2_\infty}*d\xi_t, \ J=\frac{1}{16\pi}\int_{S^2_\infty}*d\xi_\phi.
\end{eqnarray}
where $\xi_t$ and $\xi_\phi$ without the index denote 1-forms, {\it i.e.}, $\xi_t=g_{t\mu}dx^\mu$ and $\xi_\phi=g_{\phi\mu}dx^\mu$.
The (Kaluza-Klein) electric charge and magnetic charge for the gauge field $B_\mu$ are defined as, respectively,
\begin{eqnarray}
&&Q=\frac{1}{8\pi}\int_{S^2}\rho^3*{\cal  B}, \ P=\frac{1}{8\pi}\int_{S^2} {\cal B},
\end{eqnarray}
where $S^2$ denotes a closed two-surface surrounding the black hole and the two form field ${\cal B}$ is defined by ${\cal B}=\frac{1}{2}B_{\mu\nu}dx^\mu\wedge dx^\nu$.
Similarly, the electric charge and magnetic charge for the gauge field $A_\mu$ are defined as, respectively,
\begin{eqnarray}
&&q=\frac{1}{8\pi}\int_{S^2}\left(\rho*{\cal F}^{(4)}-\frac{2}{\sqrt{3}}A_5{\cal F}\right), \ p=\frac{1}{8\pi}\int_{S^2}\left({\cal F}^{(4)}-A_5{\cal B}\right).
\end{eqnarray}
where ${\cal F}=\frac{1}{2}F_{\mu\nu}dx^\mu\wedge dx^\nu$ and ${\cal F}^{(4)}=\frac{1}{2}F_{\mu\nu}^{(4)}dx^\mu\wedge dx^\nu$.
From the asymptotic behaviors of the metric and gauge potential, we can read off the values of the charges of the new solutions:
\begin{eqnarray}
&&M^{new}=M,\\
&&J^{new}=J,\\
&&Q^{new}=\tilde \beta^3P+\tilde\gamma^3Q,\\
&&P^{new}=\tilde\gamma^3 P-\tilde \beta^3 Q,\\
&&q^{new}=\sqrt{3}\tilde\beta\tilde \gamma(\tilde\gamma Q-\tilde \beta P),\\
&&p^{new}=\sqrt{3}\tilde\beta\tilde\gamma(\tilde\beta Q+\tilde\gamma P).
\end{eqnarray}
We mention that similar transformations of charges have been obtained in a 
(different) $N=2$ supergravity \cite{BKRSW}.
The scalar charge $\Sigma_\rho$ and axion charge $\Sigma_a$ are written as
\begin{eqnarray}
\Sigma_\rho=(\tilde\gamma^2-\tilde\beta^2)\Sigma,\ 
\Sigma_a=4\tilde \beta\tilde\gamma \Sigma.
\end{eqnarray}

\subsection{Horizon}

As will be shown below, the horizons exist at the values of $r=r_\pm$ which satisfies the quadratic equation $\Delta=0$ when the parameters satisfy the inequality
\begin{eqnarray}
M^2\ge P^2+Q^2+a^2-\Sigma^2,
\end{eqnarray}
which is the necessary condition for existence of horizons.
Let us introduce the new coordinates $(t',\phi',x^{\prime 5})$ defined by
\begin{eqnarray}
&&dt=dt'+F(r)dr,\\
&&d\phi=d\phi'+G(r)dr,\\
&&dx^5=dx^{\prime5}+H(r)dr,
\end{eqnarray}
where the functions $F,H$ and $H$ are defined by
\begin{eqnarray}
&&F(r)=-\omega^0{}_\phi(r_\pm) G(r),\\
&&G(r)=\frac{a}{\Delta},\\
&&H(r)=-\tilde\gamma^3\omega^5{}_\phi(r_\pm)G(r)+\tilde\beta^3 \frac{2JP}{a\Delta}\left[\frac{2Q^2}{\left(M+\frac{\Sigma}{\sqrt{3}}\right)^2-Q^2}-\frac{\left(M+\frac{\Sigma}{\sqrt{3}}\right)r_++P^2-Q^2+\Sigma\left(\sqrt{3}M-\frac{\Sigma}{3}\right)}{\left(M-\frac{\Sigma}{\sqrt{3}}\right)^2-P^2}\right].
\end{eqnarray}
Here, we denote $\omega^0{}_\phi(r_+,\theta)$ and $\omega^5{}_\phi(r_+,\theta)$ by $\omega^0{}_\phi(r_+)$ and $\omega^5{}_\phi(r_+)$ simply, respectively.
In terms of the coordinates $(t',\phi',x^{\prime 5})$, the metric can be written as
\begin{eqnarray}
ds^2&=&\rho^2\Biggl[dx^{\prime 5}+B_\mu dx^{\prime \mu}+\tilde \gamma^3 \frac{a}{f^2}\Biggl\{ 2P\cos\theta-\frac{2QJ(M-\Sigma/\sqrt{3})\sin^2\theta}{[(M+\Sigma/\sqrt{3})^2-Q^2](r-r_\mp)}-\omega^5{}_\phi(r_\pm)\nonumber\\
&&+\frac{C}{B}\left(\frac{2J\sin^2\theta}{r-r_\mp}-\omega^0{}_\phi(r_\pm)\right) \Biggr\}dr
+\tilde\beta^3\Biggl\{\frac{2JP(M+\Sigma/\sqrt{3})}{a[(M-\Sigma/\sqrt{3})^2-P^2](r-r_\mp)}\nonumber\\
&&-\frac{a(c_2(r-r_\pm)^2+c_1(r-r_\pm)+c_0)+2Qa((r-r_\pm)+d_0)\cos\theta}{(r-r_\mp)A}\Biggr\}dr\Biggr]^2+\rho^{-1}ds_{(4)}^2,
\end{eqnarray}
where the four-dimensional metric is
\begin{eqnarray}
ds_{(4)}^2&=&-\frac{f^2}{\sqrt{AB}}\left[dt'+\omega^0{_\phi d\phi'+\frac{a}{f^2}}\left(\frac{2J\sin^2\theta}{r-r_\mp}-\omega^0{}_\phi(r_\pm)\right)dr\right]^2\nonumber\\
          &+&\sqrt{AB}\left(\frac{dr^2}{f^2}+\frac{2a\sin^2\theta}{f^2}drd\phi'+d\theta^2+\frac{\Delta \sin^2\theta}{f^2}d\phi^{\prime 2}\right).
\end{eqnarray}
The constants $c_i\ (i=0,1,2)$ and $d_0$ are
\begin{eqnarray}
c_0&=&\frac{2JP}{a^2}\frac{\left(M+\frac{\Sigma}{\sqrt{3}}\right)\left(a^2-\frac{2P^2\Sigma}{\Sigma-\sqrt{3}M}+\left(r_\pm-\frac{\Sigma}{3}\right)\left(3r_\pm-\frac{\Sigma}{\sqrt{3}}\right)\right)+2\left(r_\pm-\frac{\Sigma}{\sqrt{3}}\right)(P^2-Q^2+\Sigma(\sqrt{3}M-\frac{\Sigma}{3}))}{\left(M-\frac{\Sigma }{\sqrt{3}}\right)^2-P^2}\nonumber\\
&&-\frac{4JP}{a^2}\frac{\left(M+\frac{\Sigma}{\sqrt{3}}\right)(M^2-P^2-Q^2+\Sigma^2)+2Q^2(r_\pm-\frac{\Sigma}{\sqrt{3}})}{\left(M+\frac{\Sigma}{\sqrt{3}}\right)^2-Q^2}-\frac{4JP}{a^2}(r_\pm-M),
\end{eqnarray}
\begin{eqnarray}
c_1&=&\frac{2JP}{a^2}\Biggl[\frac{3\left(M+\frac{\Sigma}{\sqrt{3}}\right)r_\pm+P^2-Q^2+\frac{\Sigma}{\sqrt{3}}(M-\sqrt{3}\Sigma)}{\left(M-\frac{\Sigma }{\sqrt{3}}\right)^2-P^2}-\frac{2 Q^2}{\left(M+\frac{\Sigma }{\sqrt{3}}\right)^2-Q^2}\Biggr],
\end{eqnarray}
\begin{eqnarray}
c_2=\frac{2 J P \left(M+\frac{\Sigma }{\sqrt{3}}\right)}{a^2 \left[\left(M-\frac{\Sigma }{\sqrt{3}}\right)^2-P^2\right]},
\end{eqnarray}
\begin{eqnarray}
d_0=2 r_\pm-\frac{2 \Sigma }{\sqrt{3}}+\frac{2 J^2 P^2 \left(M+\frac{\Sigma }{\sqrt{3}}\right)}{a^2 \left[\left(M-\frac{\Sigma }{\sqrt{3}}\right)^2-P^2\right] \left[\left(M+\frac{\Sigma }{\sqrt{3}}\right)^2-Q^2\right]}.
\end{eqnarray}
We can see that the metric well behaves at $r=r_\pm$.
\medskip
Furthermore, to see that the surfaces $r=r_\pm$ are Killing horizons, we introduce coordinates $(v,\phi^{\prime\prime},x^{\prime\prime 5})$ defined by
\begin{eqnarray}
&&dt'=dv,\\
&&d\phi=d\phi^{\prime\prime}-\omega^0{}_\phi(r_\pm)^{-1}dv,\\
&&dx^{\prime 5}=dx^{\prime\prime 5}+[(B_\phi(r_\pm)\omega^0{}_\phi(r_\pm)^{-1}-B_t(r_\pm))]dv-B_\phi(r_\pm)d\phi^{\prime\prime}.
\end{eqnarray}
Then, near $r=r_\pm$, the metric behaves as
\begin{eqnarray}
ds^2&\simeq& \rho(r_\pm)^2\left[dx^{\prime\prime 5}+f_1dr\right]^2+\rho(r_\pm)^{-1}\nonumber\\
    &\times&\left[ \frac{a^2\sin^2\theta}{\sqrt{A(r_\pm)B(r_\pm)}}\left(\omega^0{}_\phi(r_\pm)d\phi^{\prime\prime}+f_2 dr\right)^2+\sqrt{A(r_\pm)B(r_\pm)}\left(\frac{dr^2}{f(r_\pm)^2}+d\theta^2-\frac{2}{a}dr(d\phi^{\prime\prime}-\omega^0{}_\phi(r_\pm)^{-1}dv)\right) \right],\nonumber \\
\end{eqnarray}
where the functions $f_1$ and $f_2$ are defined by
\begin{eqnarray}
f_1&=&\tilde \gamma^3 \frac{a}{f(r_\pm)^2}\Biggl\{ 2P\cos\theta-\frac{2QJ(M-\Sigma/\sqrt{3})\sin^2\theta}{[(M+\Sigma/\sqrt{3})^2-Q^2](r_\pm-r_\mp)}-\omega^5{}_\phi(r_\pm)\nonumber\\
&&+\frac{C(r_\pm)}{B(r_\pm)}\left(\frac{2J\sin^2\theta}{r_\pm-r_\mp}-\omega^0{}_\phi(r_\pm)\right) \Biggr\}+\tilde\beta^3\Biggl\{\frac{2JP(M+\Sigma/\sqrt{3})}{a[(M-\Sigma/\sqrt{3})^2-P^2](r_\pm-r_\mp)}-\frac{a(c_0+2Qd_0\cos\theta)}{(r_\mp-r_\pm)A(r_\pm)}\Biggr\},\\
f_2&=&\frac{a}{f(r_\pm)^2}\left(\frac{2J\sin^2\theta}{r_\pm-r_\mp}-\omega^0{}_\phi(r_\pm)\right).
\end{eqnarray}
From this, one see that the Killing vector $V=\partial/\partial v$ becomes null on the null surfaces $r=r_\pm$. 
Moreover $V_\mu dx^\mu=g_{vr}dr$ holds there, so that $V$ is tangent to the surfaces.
Therefore this shows that the surfaces $r=r_\pm$ are Killing horizons.

\medskip
We now consider what the topology of the spatial cross section of the horizon, which is given by the roots of $\Delta=0$, is from the five-dimensional point of view. 
From the four dimensional point of view, it is evident that each $t,r$=constant surface is topologically $S^2$.
It hence follows that in the five-dimensional spacetime, each $t,r$=constant surfaces can be regarded as a $U(1)$ principle bundle over $S^2$.
In this case, to know the topology of the fiber bundle, it is convenient to consider the first Chern number of the surface.
which is defined as
\begin{eqnarray}
c_1({\cal B})=-\frac{1}{\Delta x^{5}}\int_{S^2}{\cal B},
\end{eqnarray}
where ${\cal B}=\frac{1}{2}B_{\mu\nu} dx^\mu\wedge dx^\nu$ and the periodicity of the fifth coordinate $x^5$ is set to be
\begin{eqnarray}
\Delta x^5=4\pi(2\tilde\gamma^3P-2\tilde \beta^3Q).
\end{eqnarray}
From Eq.~(\ref{eq:newmetric}), we can compute the first Chern number as
\begin{eqnarray}
\left|c_1({\cal B})\right|=1,
\end{eqnarray}
which means that each $t,r$=constant surface (hence the spatial cross section of the horizon) is diffeomorphic to $S^3$.

\section{summary and discussion}
In this paper, we have presented new Kaluza-Klein black hole solutions in 
five-dimensional minimal supergravity.
We have used the $SL(2,R)$ duality symmetry that the reduced Lagrangian possesses upon reduction to four 
dimensions.
We have also studied regularity on horizons and have shown that our black hole solutions have six conserved charges 
(but see below).  
From a four-dimensional point of view, our solutions can be regarded as dyonic rotating black holes which have 
electric and magnetic charges of the Maxwell fields in addition to 
Kaluza-Klein electric and magnetic monopole charges.
From the five-dimensional point of view, like known Kaluza-Klein charged black hole solutions, the black hole spacetime has two horizons, the outer and inner horizons, and although the cross-section geometry of the outer horizon is of $S^3$, at large distances the spacetime behaves effectively as a four-dimensional spacetime. 
 
\medskip
First, we would like to comment on how many independent parameters our solutions have. 
As mentioned previously, although our solutions carry six charges, we find that not all are independent. 
To see this, it is straightforward to compute the Jacobian, which is computed as
\begin{eqnarray}
\left|\frac{\partial(Q^{new},P^{new},q^{new},p^{new})}{\partial (Q,P,\alpha,\beta)}\right|=0. \nonumber
\end{eqnarray}
This means that the relation between the four charges $(Q^{new},P^{new},q^{new},p^{new})$ and the four parameters $(Q,P,\alpha,\beta)$ is not one-to-one. 
This fact can also be recognized as follows: In terms of $\beta$ and $\gamma(\equiv 1+\alpha\beta)$, the new charges are written as
\newcommand{\Pn}{P^{new}}
\newcommand{\Qn}{Q^{new}}
\newcommand{\pn}{p^{new}}
\newcommand{\qn}{q^{new}}
\beqa
\left(
\begin{array}{c}
\Pn\\
\Qn\\
\pn\\
\qn
\end{array}
\right)
&=&
\frac1{\beta^2+\gamma^2}
\left(
\begin{array}{cc}
\gamma^3&-\beta^3\\
\beta^3&\gamma^3\\
\beta\gamma^2&\gamma\beta^2\\
-\gamma\beta^2&\beta\gamma^2
\end{array}
\right)
\left(
\begin{array}{c}
P\\
Q
\end{array}
\right),
\eeqa
where for simplicity we divide $(p^{new},q^{new})$ by $\sqrt{3}$. 
The set of charges $\left(
\begin{array}{c}
P\\
Q
\end{array}
\right)
$ can be solved in two ways, using $x\equiv \beta/\gamma$:
\beqa
\left(
\begin{array}{c}
P\\
Q
\end{array}
\right)
&=&\frac{x^2 +1}{\gamma(x^6 +1)}\left(
\begin{array}{cc}1&x^3 \\ -x^3 &1\end{array}
\right)
\left(
\begin{array}{c}
\Pn\\
\Qn
\end{array}
\right),\\
\left(
\begin{array}{c}
P\\
Q
\end{array}
\right)
&=&\frac{1}{\gamma}\left(
\begin{array}{cc}x^{-1}&-1 \\ 1&x^{-1}\end{array}
\right)
\left(
\begin{array}{c}
\pn\\
\qn
\end{array}
\right).
\eeqa
Eliminating $\left(
\begin{array}{c}
P\\
Q
\end{array}
\right)
$ from these equations, we find
\beqa
x&=&-s\pm\sqrt{s^2+1},
\eeqa
where 
\beqa
s&\equiv&\frac12\frac{\Pn \pn + \Qn \qn}{\Qn\pn - \Pn \qn}.
\eeqa
$s$ is determined by the two ratios $\pn/\qn$ and $\Pn/\Qn$.
If the length of $\left(
\begin{array}{c}
\pn\\
\qn
\end{array}
\right)
=\sqrt{{\pn}^2+{\qn}^2}$ is fixed to some value, then $\left(
\begin{array}{c}
\pn\\
\qn
\end{array}
\right)
$
is determined, and $\left(
\begin{array}{c}
\Pn\\
\Qn
\end{array}
\right)
$
is also determined by the equation
\beqa
\left(
\begin{array}{c}
\Pn\\
\Qn
\end{array}
\right)
&=&
\left(
\begin{array}{cc}
2s&-1\\
1&2s
\end{array}
\right)
\left(
\begin{array}{c}
\pn\\
\qn
\end{array}
\right).
\label{charge_constraint}
\eeqa
Only the solutions with charges satisfying (\ref{charge_constraint}) can be generated. For this reason, our solutions do not have a limit to the Ishihara-Matsuno solutions. 

\medskip
The constraint between the charges may be regarded as a consequence of the fact that the $E_{7(+7)}$ 
U-duality transformation does not change the quartic invariant \cite{CJ}, which is made of an $E_7$ multiplet 
in the {\bf 56} representation. Indeed, by using the consistent truncation \cite{PT}, one can see what 
becomes of the quartic invariant of $E_{7(+7)}$ in the dimensionally reduced $D=5$ minimal supergravity.
The result is a certain quartic invariant of $SL(2,{\bf R})$ made out of, in this case, an $SL(2,{\bf R})$ 
multiplet in the {\bf 4} representation.

\medskip
It is natural to think that the most general black hole solutions within the theory ---if exist---  should have {\sl independent} six charges. 
In principle, our technique would allows to generate such solutions with  maximal number of parameters (a mass, an angular momentum, two electric charges, and two magnetic charges), depending on the choice of a seed and a Killing vector. 
We leave this issue open for future work.

\medskip
Moreover, in connection with  the above discussion concerning the parameter independence, we here mention the difference between the five-parameter solutions constructed in this paper and other five-parameter solutions in Ref.~\cite{GS2}, which have recently been constructed by a non-linear sigma model approach ($SO(4,4)$-duality) in five-dimensional $U(1)^3$ ungauged supergravity (as well known, identifying three $U(1)$ gauge fields and freezing out the moduli fields~\cite{GS1} in the theory yields five-dimensional minimal supergravity.). 
 For the solutions in Ref.~\cite{GS2} the dimensionally reduced solutions have not only the electric charge $q$ and magnetic charge $p$ but also a nut parameter, so that the four-dimensional metric is no longer asymptotically Minkowskian. 
When one takes the vanishing limit of the nut parameter, both the charges must vanish.    
By contrast, our solutions does not have such a nut parameter, and it turns out that when the parameters satisfy $\tilde \gamma Q=\tilde \beta P$ (or $\tilde \gamma P=-\tilde \beta Q$), the electric charge $q^{new}$ ($p^{new}$) vanishes but the magnetic charge $p^{new}$ ($q^{new}$) does not.

\medskip
Next, we provide some further remarks on the solution-generation technique by the ($D=3$)
non-linear sigma in~\cite{BCCGSW} model and the $SL(2,R)$-duality that we have used in this paper. 
Our transformation requires at least a single Killing vector, while the ($D=3$) non-linear sigma model approach 
needs the existence of two commuting Killing vectors. 
For the non-linear sigma model approach, the bosonic sector of five-dimensional minimal supergravity theory with two commuting independent
Killing symmetries can be described by three-dimensional gravity coupled with eight scalar fields $\Phi^A$ $(A=1,\cdots,8)$ describing the coset space $G_{2(+2)}/[SL(2,R)\times SL(2,R)]$.
Thus, the solutions of the original system can be expressed by 
a symmetric, unimodular 
matrix $M(\Phi^A)$. In this case, 
one must identify the target space coordinates $\Phi^A$ corresponding to the seed solution and form the coset matrix $M(\Phi^A)$ on the symmetric space $G_{2(+2)}/[SL(2,R)\times SL(2,R)]$. 
This {\sl dualization} involves solving certain linear partial derivative equations on the three-dimensional base space. 
Then, we make a desirable type of a global $G_{2(+2)}$ transformation $M(\Phi^A)\to M'(\Phi^A)=\Lambda^t M(\Phi^A)\Lambda $ for $\Lambda\in G_{2(+2)}$. 
From the relation $M(\Phi'{}^A)=M'(\Phi^A)$, one can express the new target space variables $\Phi'{}^A$ in terms of the old target space variables $\Phi^A$. 
Finally, one identifies the metric and gauge potential of the new solution with the new target space coordinates $\Phi'{}^A$, which is so-called {\sl inverse dualization} and also requires solving some linear partial differential equations. 
Although the construction of the new target space variables via the hidden symmetry transformations is a relatively simple problem, some complications often arise in solving the dualization equations for the one-form fields to scalars. 
However, for our formalism, we would like to emphasize that unlike the sigma model approach, it is only when one computes the $1$-form $\tilde B$ from the Kaluza-Klein gauge field $B$ for a seed solution in Eq.(\ref{eq:tildeB}) that one has to solve such partial differential equations.
An interesting application can be expected for the construction of a wider class of black hole solutions.

\medskip
Finally, we would like to make a few comments on the transformation preserving asymptotics. 
All the actions of the $SL(2,R)$ group on a five-dimensional Kaluza-Klein type of solution preserve its asymptotic structure, i.e., 
if we start with an asymptotically Kaluza-Klein solution as a seed, the transformed solution is also sure to be an asymptotically Kaluza-Klein solution. 
(This should be compared with the $D=3$ duality, where only the denominator subgroup $H$ of the $G/H$ 
sigma model preserves the asymptotics \cite{BMG}; the $D=4$ $SL(2,R)$ here becomes a part of the 
denominator subgroup upon further reduction to three dimensions.)
We can see this by the following simple consideration: 
When we choose the generator of the fifth direction $\partial/\partial x^5$ as a Killing vector in dimensional reduction, it is evident that for the starting-point Kaluza-Klein black hole solution, the dilaton field $\rho$ and axion field $A_5$ at infinity behave as constants. 
From Eq.~(\ref{eq:newrho}), it hence follows that under the global $SL(2,R)$ linear transformation,  the transformed dilaton field at infinity also have to behave as constant, and that furthermore under the transformation, the four-dimensional metric $ds_{(4)}^2$ is invariant.  
Therefore,  
Kaluza-Klein asymptotics is preserved by this transformation, 
while, as is easily seen, the $SL(2,R)$-transformation in general cannot preserve asymptotic flatness (asymptotic Minkowskian).

\section*{Acknowledgments} 
We would like to thank H. Kodama for valuable discussions and comments.
The work of S.~M. and S.~T. is supported by 
Grant-in-Aid
for Scientific Research  
(A) No.22244030, and S.~M. 
is also by (C) No.20540287  
from
The Ministry of Education, Culture, Sports, Science
and Technology of Japan.


\begin{thebibliography}{99}






\bibitem{Sch}
K. Schwarzschild, {\it Sitzber. Deut. Akad. Wiss. Berlin}, KL. Math.-Phys. Tech. 189 (1916).

\bibitem{p-brane} 
  M.~J.~Duff, R.~R.~Khuri and J.~X.~Lu, Phys.\ Rept.\  {\bf 259}, 213 (1995).

\bibitem{SV}
 A.~Strominger and C.~Vafa,
  Phys.\ Lett.\  B {\bf 379}, 99 (1996).

\bibitem{IM}
H. Ishihara and K. Matsuno, Prog. Theor. Phys. {\bf 116}, 417 (2006).
\bibitem{Wang}
T. Wang, Nucl. Phys. B {\bf 756}, 86 (2006). 
\bibitem{NIMT}
T. Nakagawa, H. Ishihara, K. Matsuno and S. Tomizawa, Phys. Rev. D {\bf 77}, 044040 (2008). 
\bibitem{TIMN}
S. Tomizawa, H. Ishihara, K. Matsuno and T. Nakagawa, Prog. Theor. Phys. {\bf 121}, 823 (2009).
\bibitem{TI}
S. Tomizawa and A. Ishibashi, Class. Quant. Grav. {\bf 25}, 245007 (2008).
\bibitem{T}
S. Tomizawa, e-Print: arXiv:1009.3568 [hep-th].
\bibitem{Gauntlett0}
J. P. Gauntlett, J. B. Gutowski, C. M. Hull, S. Pakis and H. S. Reall, Class. Quant. Grav. {\bf 20}, 4587 (2003). 
\bibitem{Gaiotto}
D. Gaiotto, A. Strominger and X. Yin, JHEP {\bf 02}, 023 (2006).



\bibitem{Elvang3}
H. Elvang, R. Emparan, D. Mateos and H. S. Reall, JHEP {\bf 08}, 042 (2005).
\bibitem{BKW}
I. Bena, P. Kraus and N.P. Warner, Phys. Rev. D {\bf 72}, 084019 (2005). 
\bibitem{Bena}
I. Bena and P. Kraus, Phys. Rev. D {\bf 70}, 046003 (2004). 
\bibitem{Bena3}
I. Bena and N. P. Warner, Adv. Theor. Math. Phys. {\bf 9}, 667 (2005).
\bibitem{BGRW}
I. Bena, S. Giusto, C. Ruef and N. P. Warner, JHEP {\bf 11}, 032 (2009).
\bibitem{EEMR2}
H. Elvang, R. Emparan, D. Mateos and H. S. Reall, Phys. Rev. D {\bf 71}, 024033 (2005).
\bibitem{FGPS}
J. Ford, S. Giusto, A. Peet and A. Saxena, Class. Quant. Grav. {\bf 25}, 075014 (2008).
\bibitem{CEFGS}
J. Camps, R. Emparan, P. Figueras, S. Giustod and A. Saxena, JHEP {\bf 02}, 021 (2009).
\bibitem{GRS}
S. Giusto,  S. F. Ross and  A. Saxena, JHEP {\bf 12}, 065 (2007). 
\bibitem{CBJV}
G. Compere, S. Buyl, E. Jamsin and A. Virmani, Class. Quant. Grav. {\bf 26}, 125016 (2009).
\bibitem{Gibbons-Perry}
G. W. Gibbons and M. J. Perry, Nucl. Phys. B {\bf 248}, 629 (1984).
\bibitem{CY-KK3}
M. Cveti{\v c} and D. Youm, Nucl. Phys. B {\bf 438}, 182 (1995).
\bibitem{CY-KK4}
M. Cveti{\v c} and D. Youm, Nucl. Phys. B {\bf 453}, 259 (1995).
\bibitem{CY-KK5}
M. Cveti{\v c} and D. Youm, Phys. Rev. D {\bf 52}, 2574 (1995).
\bibitem{Nelson}
W. Nelson, Phys. Rev. D {\bf 49}, 5302 (1994). 

\bibitem{CY96}
M. Cveti\v{c} and D. Youm, Nucl. Phys. B {\bf 476}, 118 (1996).
\bibitem{Herdeiro0}
C. A. R. Herdeiro, Nucl. Phys. B {\bf 665}, 189 (2003).
\bibitem{Gimon-Hashimoto}
E. Gimon and A. Hashimoto, Phys. Rev. Lett. {\bf 91}, 021601 (2003).
\bibitem{Herdeiro}
C. A. R. Herdeiro, Class. Quant. Grav. {\bf 20}, 4891 (2003).
\bibitem{Wu}
S-Q. Wu, Phys. Rev. Lett. {\bf 100}, 121301 (2008).
\bibitem{MINT}
K. Matsuno, H. Ishihara, T. Nakagawa and S. Tomizawa, Phys. Rev. D {\bf 78}, 064016 (2008). 
\bibitem{Ehlers}
 J. Ehlers, Dissertation, Univ. Hamburg (1957).
 
 \bibitem{MM}
 R.~A.~Matzner and C.~W.~Misner,
  Phys.\ Rev.\  {\bf 154}, 1229 (1967).
  
 \bibitem{Geroch}
 R.~P.~Geroch,
  J.\ Math.\ Phys.\  {\bf 12}, 918 (1971); 
  J.\ Math.\ Phys.\  {\bf 13}, 394 (1972).
  
 \bibitem{CFS}
 E.~Cremmer, J.~Scherk and S.~Ferrara,
  Phys.\ Lett.\  B {\bf 74}, 61 (1978).
  
 
 \bibitem{CJ}
 E.~Cremmer and B.~Julia,
  Nucl.\ Phys.\  B {\bf 159}, 141 (1979).
  
  \bibitem{Cremmer}
  E. Cremmer, in: gCambridge 1980, Proceedings, Superspace and Supergravityh, eds.
S. W. Hawking and M. Roc{\v e}k (Cambridge University Press, 1981) 267.

  \bibitem{CN}
  A.~H.~Chamseddine and H.~Nicolai,
  Phys.\ Lett.\  B {\bf 96}, 89 (1980).
  
\bibitem{MO}
S.~Mizoguchi and N.~Ohta,
  Phys.\ Lett.\  B {\bf 441}, 123 (1998).
  

 \bibitem{CJS}
 E.~Cremmer, B.~Julia and J.~Scherk,
  Phys.\ Lett.\  B {\bf 76}, 409 (1978).
 
 \bibitem{HT}
 C.~M.~Hull and P.~K.~Townsend,
  Nucl.\ Phys.\  B {\bf 438}, 109 (1995).
  
 \bibitem{OP}
 N.~A.~Obers and B.~Pioline,
  Phys.\ Rept.\  {\bf 318}, 113 (1999).
 
 \bibitem{MarcusSchwarz}
 N.~Marcus and J.~H.~Schwarz,
  Nucl.\ Phys.\  B {\bf 228}, 145 (1983).
  
  \bibitem{Germar}  
 S.~Mizoguchi and G.~Schr\"oder,
  Class.\ Quant.\ Grav.\  {\bf 17}, 835 (2000).
  
  
 \bibitem{MizoguchiE10}
  S.~Mizoguchi,
  Nucl.\ Phys.\  B {\bf 528}, 238 (1998).

\bibitem{Rasheed}
D. Rasheed, Nucl. Phys. B {\bf 454}, 379 (1995). 

\bibitem{TYM}
S. Tomizawa, Y. Yasui and Y. Morisawa, Class. Quant. Grav. {\bf 26}, 145006 (2009).
\bibitem{GS1}
D. V. Gal'tsov and N. G. Scherbluk, Phys. Rev. D {\bf 78}, 064033 (2008).
\bibitem{GS2}
D. V. Gal'tsov and N. G. Scherbluk, Phys. Rev. D {\bf 79}, 064020 (2009).

\bibitem{Tomizawa}
S. Tomizawa, Phys. Rev. D {\bf 82}, 104047 (2010).








\bibitem{BCCGSW}
A. Bouchareb, G. Clement, C-M. Chen, D. V. Gal'tsov, N. G. Scherbluk and T. Wolf, Phys. Rev. D {\bf 76},104032 (2007); Erratum-ibid. D {\bf 78}, 029901 (2008).


\bibitem{Giusto-Saxena}
S. Giusto and A. Saxena, Class. Quantum Grav. {\bf 24}, 4269 (2007).


\bibitem{BMPV}
J. C. Breckenridge, R.C. Myers, A.W. Peet and C. Vafa, Phys. Lett. B {\bf 391}, 93 (1997).
\bibitem{Elvang}
H. Elvang, R. Emparan, D. Mateos and H. S. Reall,
Phys. Rev. Lett. {\bf 93}, 211302 (2004).



\bibitem{Rasheed-Gibbons1}
G. W. Gibbons and D. A. Rasheed, Nucl. Phys. B {\bf 454}, 185 (1995).
\bibitem{Rasheed-Gibbons2}
G. W. Gibbons and D. A. Rasheed, Phys. Lett. B {\bf 365}, 46 (1996).


\bibitem{PT}
  G.~Papadopoulos and P.~K.~Townsend,
  Phys.\ Lett.\  B {\bf 380}, 273 (1996)
  [arXiv:hep-th/9603087].


\bibitem{BKRSW}
K. Behrndt, R. Kallosh, J. Rahmfeld, M. Shmakova and W. K. Wong, Phys. Rev. D {\bf 54}, 6293 (1996).
 
 
\bibitem{BMG}
P. Breitenlohner, D. Maison and G. W. Gibbons, Commun. Math. Phys. {\bf 120}, 295 (1988).


\bibitem{CdBSV}
  G.~Compere, S.~de Buyl, S.~Stotyn and A.~Virmani,
  JHEP {\bf 1011}, 133 (2010)
  [arXiv:1006.5464 [hep-th]].



\end{thebibliography}
\end{document}